\documentclass[onecolumn,pre,a4paper]{revtex4}
\usepackage{graphicx}
\usepackage{amsmath}
\usepackage{psfrag}
\usepackage{latexsym}

\newcommand{\1}{\mathbf{1}}
\newcommand{\gc}{\mathbf{g_c}}
\newcommand{\gcc}{\mathbf{g_{*c}}}
\newcommand{\GC}{\mathbf{G_C}}
\newcommand{\ga}{\mathbf{g_a}}
\newcommand{\GA}{\mathbf{G_A}}
\newcommand{\Sigmacc}{\mathbf{\Sigma_{*c}}}
\newcommand{\Sigmac}{\mathbf{\Sigma_c}}

\newcommand{\A}{\mathbf{A}}
\newcommand{\X}{\mathbf{X}}

\newcommand{\ec}{\mathbf{c}}
\newcommand{\C}{\mathbf{C}}
\newcommand{\M}{\mathbf{M}}
\newcommand{\ea}{\mathbf{a}}
\newcommand{\Tr}{\mbox{Tr}}

\begin{document}

\title{\bf Spectral Moments of Correlated Wishart Matrices}
\author{Zdzis\l{}aw Burda}\thanks{burda@th.if.uj.edu.pl}
\author{Jerzy Jurkiewicz}\thanks{jjurkiew@th.if.uj.edu.pl}
\author{Bart\l{}omiej Wac\l{}aw}\thanks{bwaclaw@th.if.uj.edu.pl}

\affiliation{Mark Kac Center for Complex Systems Research and
Marian Smoluchowski Institute of Physics, \\
Jagellonian University, ul. Reymonta 4, 30-059 Krakow, Poland}

\begin{abstract}
We present an analytic method to determine  
spectral properties of the covariance matrices 
constructed of correlated Wishart random matrices. 
The method gives, in the limit of large matrices, 
exact analytic relations between the spectral moments 
and the eigenvalue densities of the covariance matrices 
and their estimators. 
The results can be used in practice 
to extract information about the genuine 
correlations from the given experimental realization 
of random matrices. 
\end{abstract}

\maketitle

Wishart random matrices play an important role in the
multivariate statistical analysis \cite{w,a}. 
They are useful in some problems
of fundamental physics \cite{dj}, 
communication and information theory \cite{mea,sm1,s},
internet trading \cite{mz} 
and quantitative finance \cite{lcbp,pea,bj}. 

A Wishart ensemble of correlated random matrices is
defined by a Gaussian probability measure:
\begin{equation}
P(\X) D\X = 
\mathcal{N}^{-1} \exp\left[ -\frac{1}{2} \Tr \,
\X^\tau \C^{-1} \X \A^{-1} \right] \prod_{i,\alpha=1}^{N,T} 
d X_{i\alpha} \ ,
\label{preal}
\end{equation}
where $\X=(X_{i\alpha})$ is a real rectangular matrix of
dimension $N\times T$. It has two types of indices:
an $N$-type index runnig over the set $i=1,\dots,N$ 
and a $T$-type index over $\alpha=1,\dots,T$. 
Throughout the paper the $N$-type indices will be denoted
by Latin letters and the $T$-type by Greek ones.
$\X^{\tau}$ denotes the transpose of $\X$.
The matrices $\C=(C_{ij})$ and $\A=(A_{\alpha\beta})$ are 
symmetric square matrices of dimensions $N\times N$ and $T\times T$, 
respectively. They are positive definite. 
$\mathcal{N}$ is a normalization constant:
\begin{equation}
\mathcal{N} = (2\pi)^\frac{NT}{2} (\det \C)^\frac{T}{2} 
(\det \A)^\frac{N}{2},
\end{equation}
chosen to have $\int P(\X) D\X = 1$. 

Let $Q(\X)$ be a quantity depending on $\X$. 
The average of $Q$ over the random matrix ensemble (\ref{preal}) 
is defined as: 
\begin{equation}
\langle Q \rangle = \int Q(\X) P(\X) D\X.
\end{equation}
In particular, the two-point correlation function is:
\begin{equation}
\langle X_{i\alpha} X_{j\beta} \rangle = C_{ij} A_{\alpha\beta},
\label{CA}
\end{equation}
as directly follows from the Gaussian integration. In this paper
we are interested in the spectral behaviour, 
the eigenvalue distribution and the spectral moments
of the following random matrices:
\begin{equation} \ec=\frac{1}{T} \X\X^{\tau} \quad , \quad
\ea=\frac{1}{N} \X^{\tau}\X \ .
\label{ca}
\end{equation}
These matrices can be used as estimators of
the correlation matrices $\C$ and $\A$ if some
realizations of random matrices $\X$ are given. We will refer to $\ec$
and $\ea$ as to covariance matrices or 
statistically dressed correlation matrices.
We will present an analytic method to determine 
the eigenvalue distribution and the spectral moments
of $\ec$ and $\ea$ in the limit of large matrix size. 
Another method of calculating the eigenvalue density of correlated
Wishart matrices has been recently discussed in \cite{sm2}.

In parallel to (\ref{preal}) one can define a Wishart 
ensemble of correlated complex matrices:
\begin{equation}
\mathcal{P}(\X) D\X = 
\mathcal{N}^{-1} \exp\left[ - \Tr \,
\X^{\dagger} \C^{-1} \X \A^{-1} \right] \prod_{i, \alpha} 
d  X^{re}_{i\alpha} d X^{im}_{i\alpha} .
\label{pcompl}
\end{equation}
The matrices $\C$ and $\A$ are now Hermitean and positive definite. 
$\X^{\dagger}$ denotes the Hermitean conjugate of $\X$.
The normalization constant is now
$\mathcal{N} = (\pi)^{NT} (\det \C)^{T} (\det \A)^{N}$.
In the analysis of the complex ensemble the estimators
(\ref{ca}) of the correlation matrices 
are replaced correspondingly by 
\begin{equation}
\ec=\frac{1}{T} \X\X^{\dagger} \quad , \quad
\ea=\frac{1}{N} \X^{\dagger}\X \ .
\label{cac}
\end{equation}
Notice that the factor one half in front of the trace
in the measure for real matrices (\ref{preal}) is dropped 
in (\ref{pcompl}). With this choice of the measure
the two-point correlations take a similar form
as for real matrices (\ref{CA}):
\begin{equation}
\langle X_{i\alpha} X^*_{j\beta}\rangle = C_{ij} A_{\alpha\beta} \ .
\label{CAc}
\end{equation}
The star stands for 
the complex conjugation. Additionally we also have:
\begin{equation}
\langle X_{i\alpha} X_{j\beta}\rangle = 
\langle X^*_{i\alpha} X^*_{j\beta}\rangle = 0 \ .
\label{CAcc}
\end{equation}
As a consequence, as we shall discuss towards the end of the paper,
the matrices $\ec$ and $\ea$ (\ref{ca}) in the real ensemble 
have an identical large $N$ behaviour 
as the corresponding matrices (\ref{cac}) in the complex ensemble.
Since we are interested here only in the large $N$
behaviour it is sufficient to consider one of the two 
ensembles and draw conclusions for the other. We will focus
the presentation on the ensemble of real matrices.

An example of a problem which can
be formulated in terms of Wishart random matrices 
(\ref{preal}) is the following.
Imagine that we probe a statistical system of $N$ correlated 
degrees of freedom by doing $T$ measurements. We store the
measured values of the $i$-th degree of freedom 
in the $\alpha$-th measurement in a rectangular
matrix $\X = (X_{i\alpha})$. The degrees of freedom
as well as the measurements may be correlated. 
This is expressed by the equation (\ref{CA}), which tells us
that the covariance matrix for the correlations between 
degrees of freedom in the system is $\C=(C_{ij})$ and for the
(auto)correlation between measurements is $\A=(A_{\alpha\beta})$.
Note that in general the correlations between $X_{i\alpha}$ 
and $X_{j\beta}$ may have a more complicated form:
$\langle X_{i\alpha} X_{j\beta} \rangle = 
\mathcal{C}_{i\alpha,j\beta}$, where the matrix $\mathcal{C}$ has 
double indices. Such a situation takes place
if the autocorrelations are different for various degrees 
of freedom. We shall not discuss this case here.
Moreover, we shall assume that only Gaussian effects are important
for the studied system.

A perfect example of the situation described above
is the problem of optimal portfolio assessment - one of
the fundamental problems of quantitative finance. The portfolio
assessment is based on the knowledge of the covariance matrix $\C$ 
for stocks returns \cite{ret}.
In practice, the covariance matrix is estimated from the historical data
which are stored in a rectangular matrix representing
$T$ historical values of $N$ stocks.
Fluctuations of returns are well described by the Gaussian 
ensemble (\ref{preal}).
The estimator of the covariance matrix is given by (\ref{ca}).
Another problem of modern financial analysis 
which can be directly cast into the form (\ref{preal}) is
the problem of taste matching \cite{mz}. This problem is encountered
for instance in the large-scale internet trading.

It is worth mentioning that the
random matrix framework may also be used in a statistical description
of data generated in Monte Carlo simulations 
for a system with many degrees of freedom, in particular of data
concerning the correlation functions. One frequently
encounters such a problem in Monte-Carlo simulations of
lattice field theory, where the field is represented
by correlated numbers distributed on a lattice. Usually, 
one is forced to use a dynamical Monte Carlo algorithm 
to sample such a system. The basic idea standing behind 
a dynamical algorithm is to 
generate a Markov chain -- a sort of a
random walk -- in the space of configurations. The degrees
of freedom on the lattice as well as the successive
configurations are usually correlated. Outside a critical region 
no long range correlations are observed and the fluctuations 
can be treated as Gaussian. 
 
Complex random matrices are useful
for instance in telecommunication  
or information theory \cite{mea,sm1,s}.

Let us come back to the ensemble of real matrices (\ref{preal}).
As we mentioned the matrices (\ref{ca}) can be treated as estimators
of the correlation matrices $\C$ and $\A$. Indeed,
from equation (\ref{CA}) we see that:
\begin{eqnarray}
\left\langle c_{ij} \right\rangle = 
\left\langle \frac{1}{T} \sum_{\alpha} X_{i\alpha} X_{j\alpha} 
\right\rangle = M_{{\A}1} C_{ij}, \label{cest} \\
\left\langle a_{\alpha\beta} \right\rangle = 
\left\langle \frac{1}{N} \sum_{i} X_{i\alpha} X_{i\beta} 
\right\rangle = M_{{\C}1} A_{\alpha\beta}, \label{aest}
\end{eqnarray}
where $M_{{\C}1} = \frac{1}{N} \Tr\; \C$ 
and $M_{{\A}1} = \frac{1}{T} \Tr\; \A$. 
This notation will be explained later. 
The last equation tells us that measuring the average of
$\ec$ over the ensemble (\ref{preal})
we obtain the matrix $\C$ up to a constant. In other words
having a realization of random matrices $\X$ we can use
(\ref{ca}) to estimate $\C$.
Similarly, we can use $\ea$ to estimate $\A$. 
Notice that the measure (\ref{preal}) is invariant under 
the transformation $\C\rightarrow b \, \C$ and 
$\A \rightarrow b^{-1} \A$, where $b$ is an
arbitrary positive real number. In particular 
$\langle c_{ij} \rangle $ and 
$\langle a_{\alpha\beta} \rangle $ are independent of the
rescaling factor $b$. This independence is ensured by the
presence of the factors $M_{{\A}1}$ and $M_{{\C}1}$ in 
equations (\ref{cest},\ref{aest}). In practical calculations,
if  $\Tr \, \A$ and $\Tr \, \C$ are not specified, 
one can remove the redundancy with respect to the rescaling
by $b$, setting $\frac{1}{T} \Tr \, \A = \frac{1}{N} \Tr \, \C$. 
In this case the constants $M_{{\A}1}=M_{{\C}1}$ can be determined 
from the data by evaluating the traces of $\ec$ or $\ea$:	 
\begin{eqnarray}
M_{{\C}1} = M_{{\A}1} = \sqrt{\frac{1}{N} \Tr \; 
\langle \ec \rangle} = 
\sqrt{\frac{1}{T} \Tr \; \langle \ea \rangle}  \;\; .
\nonumber
\end{eqnarray}

While considering the covariance matrices for
the Wishart ensemble we can formulate two reciprocal problems,
which we shall call {\bf direct} and {\bf inverse problem}.
In the direct problem 
we want to learn as much as possible about 
the probability distribution of the estimators $\ec$ and $\ea$ (\ref{ca})
assuming that the matrices $\C$ and $\A$ are given.
In particular, we want to calculate the eigenvalue density
functions: 
\begin{eqnarray}
&& \rho_{\ec}(\lambda) = 
\left\langle \frac{1}{N} 
\sum_{i=1}^N \delta(\lambda - \lambda_i) \right\rangle \ , \nonumber \\
&& \rho_{\ea}(\lambda) = 
\left\langle \frac{1}{T} 
\sum_{\alpha=1}^T \delta(\lambda - \lambda_\alpha) \right\rangle \ ,
\nonumber
\end{eqnarray}
where $\lambda_i$ and $\lambda_\alpha$ are eigenvalues of $\ec$ and $\ea$,
respectively. The determination of the eigenvalue density functions
is equivalent to the determination of all their spectral moments:
\begin{eqnarray}
&& m_{\ec k} = \int d\lambda \, \rho_{\ec}(\lambda) \, \lambda^k  = 
\left\langle \frac{1}{N} \Tr \, \ec^k \right\rangle \ , \nonumber \\
&& m_{\ea k} = \int d\lambda \, \rho_{\ea}(\lambda) \, \lambda^k = 
\left\langle \frac{1}{T} \Tr \, \ea^k \right\rangle \ . \nonumber
\end{eqnarray}
The moments $m_{\ec k},m_{\ea k}$ are related to each other:
\begin{equation}
m_{\ea k} = r^{1-k} m_{\ec k} \ ,
\label{mak_mck}
\end{equation}
where $r=N/T$, as follows from the cyclicity of the trace:
\begin{eqnarray}
{\rm Tr} \; \X\X^\tau \cdots \X\X^\tau =
{\rm Tr} \; \X^\tau \X \cdots \X^\tau \X \ . \nonumber
\end{eqnarray}
In the inverse problem we want to learn as much 
as possible about the genuine correlations in the system,
which are given by $\C$ and $\A$, 
using a measured sample of random matrices $\X$.
We can do this by computing the estimators $\ec$ and $\ea$ (\ref{ca})
and relating them to matrices $\C,\A$.
In particular we would like to estimate the eigenvalue 
distributions and the moments of $\C,\A$:
\begin{eqnarray}
&& 
M_{\C k} = \frac{1}{N} \Tr \; \C^k =
\frac{1}{N} \sum_{i=1}^N \Lambda^k_i  \ , \nonumber \\
&& M_{\A k} = \frac{1}{T} \Tr \; \A^k =
\frac{1}{T} \sum_{\alpha=1}^T \Lambda^k_\alpha \ , \nonumber 
\end{eqnarray}
where $\Lambda_i$ and $\Lambda_\alpha$ are eigenvalues 
of $\C$ and $\A$, respectively. 
The inverse problem is very important for practical applications, 
since in practice it is very common to reconstruct the properties of 
the underlying system from the experimental data.

In the analysis of the spectral properties of the matrices
$\ea$ and $\ec$ it is convenient to apply the 
Green's function technique. One can define Green's functions
for the correlation matrix $\C$ and its statistically 
fluctuating counterpart $\ec$:
\begin{eqnarray}
&& \GC (z) = \frac{1}{z\1_N-\C} \ , \\
&& \gc (z) = \left\langle \frac{1}{z\1_N-\ec} \right\rangle ,
\label{gc}
\end{eqnarray}
and correspondingly $\GA(z)$ and $\ga(z)$ for $\A$ and $\ea$. 
The symbol $\1_N$ stands for the $N\times N$ identity matrix.
A corresponding symbol $\1_T$ appears in the definition of
$\GA(z)$ and $\ga(z)$. 
The Green's functions are related to the generating functions
for the moments:
\begin{eqnarray}
&& M_{\C}(z) = \sum_{k=1}^\infty \frac{M_{\C k}}{z^k}
= \frac{1}{N} \Tr \, \bigg( z \GC(z)\bigg) - 1 \ , \label{Mcz} \\
&& m_{\ec}(z) = \sum_{k=1}^\infty \frac{m_{\ec k}}{z^k} 
= \frac{1}{N} \Tr \, \bigg( z \gc(z)\bigg) - 1 \ , \label{mcz} 
\end{eqnarray} 
or inversely:
\begin{eqnarray}
&& \frac{1}{N} \Tr \,  \GC(z)  = \frac{1+M_{\C}(z)}{z} \ , \nonumber \\
&& \frac{1}{N} \Tr \,  \gc(z)  = \frac{1+m_{\ec}(z)}{z} \ .  \label{gm}
\end{eqnarray}
The analogous relations exist for $\GA(z)$ and $\ga(z)$. 
The Green's functions can be used for finding the densities of eigenvalues:
\begin{eqnarray}
\rho_{\ec}(\lambda) = 
-\frac{1}{\pi} \mbox{Im} \, \frac{1}{N} \Tr \, \gc(\lambda+ i 0^+) = 
-\frac{1}{\pi} \mbox{Im} \, \frac{1 + m_{\ec}(\lambda+i0^+)}{\lambda+i0^+} \ ,
\label{rhoc}
\end{eqnarray}
and similarly for $\rho_{\ea}(\lambda)$. 
The eigenvalue densities $\rho_{\ea}(\lambda)$ and $\rho_{\ec}(\lambda)$
are not independent.
As follows from (\ref{mak_mck}) the corresponding generating 
functions (\ref{mcz}) fulfill the equation:
\begin{equation}
m_{\ea}(z) = r m_{\ec}(r z) \; . \label{ma=mc} 
\end{equation}
Combining the last equation with (\ref{gm}) we obtain:
\begin{equation}
\frac{1}{T} {\rm Tr} \; \ga(z) = 
r^2 \frac{1}{N} {\rm Tr} \; \gc(rz) + \frac{1-r}{z} \ .
\label{gac}
\end{equation}
Applying now (\ref{rhoc}) we have:
\begin{equation}
\rho_{\ea}(\lambda) = r^2 \rho_{\ec}(r\lambda) + (1-r) \delta(\lambda) \ .
\label{rac}
\end{equation}
The meaning of the last term on the right hand side of this equation
is that there are $T-N$ zero modes in the matrix $\ea$
if $T>N$. The zero modes disappear when $N=T$. 
Moving the term containing the delta function
to the other side of equation, dividing both sides of the
equation by $r^2$ and substituting the parameter 
$r$ by $s = T/N =1/r$ we obtain:
\begin{equation}
\rho_{\ec}(\lambda) = s^2 \rho_{\ea}(s\lambda) + (1-s) \delta(\lambda) \ .
\label{rca}
\end{equation}
Therefore, for $T<N$ the zero modes
appear in the spectrum $\rho_{\ec}(\lambda)$. 
In this case it is more convenient to use the 
parameter $s=1/r$ instead of $r$. 
The zero modes appear in the 
eigenvalue distribution of either $\ea$ or $\ec$. 
The two equations (\ref{rac}) 
and (\ref{rca}) are dual to each other.  
For $r=s=1$ they are identical. 
Because of the duality it is sufficient to solve the problem
for $r\le 1$. We will present a solution for the limit 
$r=N/T=\mbox{const}\le 1$ and $N\rightarrow \infty$
neglecting effects of the order $1/N$.

Using a diagrammatic method \cite{fz,sm3,bgjj} one
can write down a closed set of equations for 
the Green's function $\gc(z)$ (\ref{gc}):
\begin{eqnarray}
&& \gc(z) = \frac{1}{z\1_N -\Sigmac(z)} \quad , \quad
\gcc(z) = \frac{1}{T\1_T - \Sigmacc(z)} \ , \nonumber \\ 
\label{4gc} \\
&& \Sigmac(z) = \C \mbox{Tr} \, (\A \, \gcc(z)) \quad , \quad
\Sigmacc(z) = \A \mbox{Tr} \, (\C \, \gc(z)) \ . \nonumber
\end{eqnarray}
The set contains four equation for four unknown matrices
including $\gc(z)$ which we want to calculate, and
three auxiliary ones:
$\gcc(z)$, $\Sigmac(z)$, $\Sigmacc(z)$ (see Appendix 1).
Each of them can be interpreted in terms of a generating function for 
appropriately weighted diagrams with two external lines:
$\gc(z), \gcc(z)$ for all diagrams 
and $\Sigmac(z), \Sigmacc(z)$
for one-line-irreducible diagrams \cite{fz,sm3,bgjj} (see Appendix 1). 
In the limit $N\rightarrow \infty$ the weights of
non-planar diagrams vanish at least as $1/N$.
Thus in this limit only planar diagrams give a contribution 
to the Green's function. 
Therefore the large $N$ limit is alternatively called
the planar limit. The diagrammatic equations (\ref{4gc})
hold only in this limit. An analogous set of equations
can be written for the Green's function $\ga(z)$. The equations
are identical to those of (\ref{4gc}) if one 
exchanges $\ea \longleftrightarrow \ec$, $\A \longleftrightarrow \C$ 
and $T \longleftrightarrow N$. The two sets can be solved 
independently of each other. However, as follows 
from the duality (\ref{gac}) it is sufficient to solve
only one of them and deduce the solution of the other.

The equations (\ref{4gc}) can be solved for $\gc(z)$
by a successive elimination of 
$\gcc(z)$, $\Sigmac(z)$ and $\Sigmacc(z)$. 
However, the resulting equation is very entangled \cite{sm3}:
\begin{equation}
\gc(z) =
\left( z\1_N - \C {\rm Tr} \; 
\frac{\A}{\1_T T- \A {\rm Tr}\; (\C \gc(z))} \right)^{-1},
\label{compl}
\end{equation}
and cannot be easily used in practical calculations
of the moments $m_{\ec k}$ or spectral density $\rho_{\ec}(\lambda)$.	
 
Another way of solving the equations (\ref{4gc})
was proposed in \cite{bgjj}.
It relies on introducing a new complex variable 
$Z$ conjugate to $z$ which is defined by the equation:
\begin{equation}
m_{\ec}(z) = M_{\C}(Z) \ .
\label{map}
\end{equation} 
At the first glance this equation looks useless because
it refers to an unknown function $m_{\ec}(z)$ which we actually 
want to determine. Quite contrary to this,
as we shall see, the introduction of the conjugate 
variable $Z$ allows us to write down
a closed functional equation for $m_{\ec}(z)$.
First, let us illustrate how the method works 
for $\A=\1_T$ \cite{bgjj}. 
In this case, the elimination of the auxiliary functions
(\ref{4gc}) leads to 
\begin{equation}
Z = \frac{z}{1 + r m_{\ec}(z)} \ ,
\label{Zz}
\end{equation}
or equivalently to
\begin{equation}
z = Z \cdot (1 + r M_{\C}(Z)) \ .
\label{zZ}
\end{equation}
Suppose we solve the direct problem. In this case 
we know the matrix $\C$ and hence also the generating function
$M_{\C}(Z)$. Inserting (\ref{Zz}) to (\ref{map}) 
we obtain a closed compact functional relation  
for $m_{\ec}(z)$:
\begin{equation}
m_{\ec}(z) = M_{\C}\left(\frac{z}{1+r m_{\ec}(z)}\right).
\label{mcMC}
\end{equation}
If $M_{\C}(z)$ has a simple form,
one can solve the equation for $m_{\ec}(z)$ analytically \cite{bgjj}. 
In general, one can write a numerical program to calculate
the eigenvalue density $\rho_{\ec}(\lambda)$ from the last
equation. In case of solving the inverse problem, we assume
that we can determine moments $m_{{\ec}k}$ from the data
and hence that we can approximate the generating
function $m_{\ec}(z)$. Then we can insert (\ref{zZ}) 
to (\ref{map}) and obtain a functional
equation for $M_{\C}(Z)$: 
\begin{equation}
M_{\C}(Z) = m_{\ec}\left(Z \cdot (1 + r M_{\C}(Z))\right).
\label{MCmc}
\end{equation}
The problem is solved in principle. However, in practical
terms the inverse problem is much more difficult, because one 
cannot compute all experimental moments $m_{{\ec}k}$ with an
arbitrary accuracy, unless one has an infinitely long series 
of measurements. But one never has.
In practice one can estimate only a few lower moments 
$m_{{\ec}k}$ with a good accuracy. Because of this practical
limitation one cannot entirely solve the inverse problem.
However, as we discussed in \cite{bj} the inverse problem
can be partially solved even in specific practical applications
using a moments method. Let us sketch this method below.

We can gain some insight into the spectral
properties of the correlation matrix $\C$ by determining
the relation between the moments $M_{{\C}k}$ and $m_{{\ec}k}$.
Expanding the functions $M_{\C}(z)$ and $m_{\ec}(z)$ in (\ref{mcMC})
in $1/z$ using (\ref{Mcz},\ref{mcz}) and comparing
the coefficients at $1/z^k$ we obtain:
\begin{equation}
\begin{array}{rl}
m_{{\ec}1} &= M_{{\C}1} \\
m_{{\ec}2} &= M_{{\C}2} + r M_{{\C}1}^2 \\
m_{{\ec}3} &= M_{{\C}3} + 3 r M_{{\C}1} M_{{\C}2} + r^2 M_{{\C}1}^3 \\
m_{{\ec}4} &= M_{{\C}4} + 2 r \left(M_{{\C}2}^2 + 
2 M_{{\C}1} M_{{\C}3}\right)
+ 6 r^2 M_{{\C}1}^2 M_{{\C}2} + r^3 M_{{\C}1}^4 \\
       &\dots 
\label{mM}
\end{array}
\end{equation}
We can also invert the equations for $M_{{\C}k}$.
The result of inversion gives a set of equations which
can be directly obtained from the $1/Z$ expansion of the functions
in the equation (\ref{MCmc}) 
which is the inverse transform of (\ref{mcMC}).
We can also determine the corresponding relations
for 'negative' moments $m_{{\ec}k}$ and $M_{{\C}k}$, 
that is for $k<0$, or determine the spectral 
density $\rho_{\ec}(\lambda)$ \cite{bgjj}.
Using a computer tool for symbolic calculations 
one can easily write a program which successively 
generates the relations between spectral moments (\ref{mM})
from the equation (\ref{mcMC}).

The calculations get more complicated in the general
case when both ${\C}$ and ${\A}$ are arbitrary.  
The guiding principle is the same, though. We introduce the conjugate
variable $Z$ (\ref{map}) and, using it, write down
the solution of the equations (\ref{4gc}).
In the direct problem we assume that
the generating functions $M_{\C}(Z)$ and $M_{\A}(Z)$ are known.
We will show that in this case the solution of (\ref{4gc})
takes a form of an explicit equation for $z=z(Z)$, where the function
$z(Z)$ depends on the functions $M_{\C}(Z)$ and $M_{\A}(Z)$. Inserting
this solution back to (\ref{4gc}) we eventually obtain
a functional equation $m_{\ec}(z(Z))=M_{\C}(Z)$ from which we can 
extract the function $m_{\ec}(z)$. 

The solution of (\ref{4gc}) takes the form:
\begin{equation}
\frac{z}{Z} = \frac{1}{T} \sum_{\alpha=1}^T 
\frac{\Lambda_{\alpha}}{1- \Lambda_{\alpha} r \frac{Z}{z} M_{\C}(Z)},
\label{zZsum}
\end{equation}
where $\Lambda_\alpha$ are eigenvalues of ${\A}$. It can 
be rewritten as:
\begin{equation}
r M_{\C}(Z) = M_{\A}\left(\frac{z}{rZM_{\C}(Z)}\right) \ .
\end{equation}
and can be formally solved for $z$:
\begin{equation}
z =  Z \, rM_{\C}(Z) \, M_{\A}^{-1} \left(rM_{\C}(Z)\right),
\label{zZ_CA}
\end{equation}
where $M_{\A}^{-1}$ is the inverse function of $M_{\A}$.
Thus we have obtained an explicit equation for
$z=z(Z)$ in terms of the known functions $M_{\A}$ and $M_{\C}$.
One can easily check that for ${\A}=\1_T$ the last equation 
reduces to (\ref{zZ}). In this case $M_{\A}(z) = 1/(z-1)$,
$M_{\A}^{-1}(z) = 1 + 1/z$.

Combining the equation for $z=z(Z)$ given by (\ref{zZ_CA}) with
(\ref{map}) we arrive at a closed equation for
the generating function $m_{\ec}(Z)$. It can be used
for example to calculate the moments 
$m_{{\ec}k}$'s (see Appendix 2). The calculations
yield a set of equations expressing $m_{{\ec}k}$'s in terms
of the bare moments $M_{{\A}k}$ and $M_{{\C}k}$:
\begin{eqnarray}
&& m_{{\ec}1}  = M_{{\C}1} M_{{\A}1}  \nonumber \\
&& m_{{\ec}2}  = M_{{\C}2} M_{{\A}1}^2  + r M_{{\C}1}^2 M_{{\A}2} 
\nonumber \\
&& m_{{\ec}3}  = M_{{\C}3} M_{{\A}1}^3  + 
                  3r M_{{\C}1} M_{{\C}2} M_{{\A}1} M_{{\A}2} 
                + r^2 M_{{\C}1}^3 M_{{\A}3}  \label{mcMM} \\
&& m_{{\ec}4}  = M_{{\C}4} M_{{\A}1}^4 + 
               2r \left(M_{{\C}2}^2 + 2 M_{{\C}1}M_{{\C}3}\right) 
                M_{{\A}1}^2 M_{{\A}2}
             + 2r^2 M_{{\C}1}^2 M_{{\C}2}\left(M_{{\A}2}^2 + 
                 2 M_{{\A}1}M_{{\A}3}\right) 
             + r^3 M_{{\C}1}^4 M_{{\A}4}  \nonumber \\
&&             \dots \nonumber
\end{eqnarray}
The equations reduce to the 
form (\ref{mM}) for ${\A}=\1_T$.

Using the relations (\ref{mak_mck}) we can also determine the
moments of the matrix $\ea$. It is more convenient to write them
using the variable $s=r^{-1}$ -- the dual counterpart of $r$ --
instead of $r$ itself:
\begin{eqnarray}
&& m_{{\ea}1}  = M_{{\A}1} M_{{\C}1}  \nonumber \\
&& m_{{\ea}2}  = M_{{\A}2} M_{{\C}1}^2  + s M_{{\A}1}^2 M_{{\C}2} 
\nonumber \\
&& m_{{\ea}3}  = M_{{\A}3} M_{{\C}1}^3  + 
                  3s M_{{\A}1} M_{{\A}2} M_{{\C}1} M_{{\C}2} 
                + s^2 M_{{\A}1}^3 M_{{\C}3}  \label{maMM} \\
&& m_{{\ea}4}  = M_{{\A}4} M_{{\C}1}^4 + 
               2s \left(M_{{\A}2}^2 + 2 M_{{\A}1}M_{{\A}3}\right) 
                 M_{{\C}1}^2 M_{{\C}2}
               + 2s^2 M_{{\A}1}^2 M_{{\A}2}
                \left(M_{{\C}2}^2 + 2 M_{{\C}1}M_{{\C}3}\right) 
             + s^3 M_{{\A}1}^4 M_{{\C}4} \nonumber \\
&&             \dots \nonumber
\end{eqnarray}
The equations are completely symmetric to (\ref{mcMM}) 
with respect to the change
$r \longleftrightarrow s$ (which amounts to $N \longleftrightarrow T$),
and ${\ec} \longleftrightarrow \ea$, ${\C} \longleftrightarrow {\A}$.
Using this method one can obtain equations (\ref{mcMM}) and (\ref{maMM})
to an arbitrary order.

The above relations are useful for computing the dressed moments 
$m_{\ea k},m_{\ec k}$ for given matrices $\A,\C$ or inversely, 
the genuine moments $M_{\A k},M_{\C k}$ from the experimental data. 
As mentioned, the spectral moments give us in principle
full information about the eigenvalue distribution. 
In practice the reconstruction of the eigenvalue
density may be difficult, because to do it
we would need to know all moments with a very good precision.
Usually, in practical applications one can accurately evaluate 
only a few lower moments. 

In some special cases if we can make some extra assumptions
about the form of the matrices $\C$ or $\A$ we can improve
significantly the reconstruction of the eigenvalue density.
In the previous work \cite{bgjj} we have analysed the case of $\A=\1_T$
and of the matrix $\C$ which had only a few distinct eigenvalues. 
In this case the Green's function $\gc(z)$ is given by an
algebraic equation of the order which is equal to the number
of distinct eigenvalues. It can be analytically solved when 
this number
is less or equal four. If it is larger the problem can be
handled numerically. The duality tells us that the solution
also holds when we change the roles of $\A$ and $\C$.

\begin{figure}
\begin{center}
\includegraphics[width=10cm]{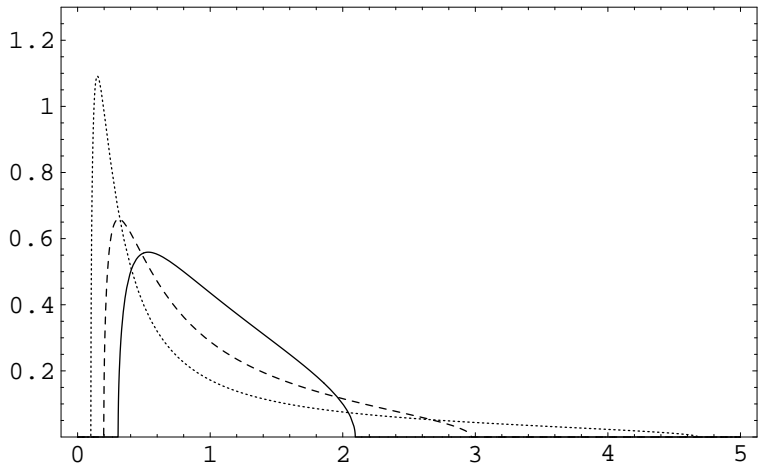}
\end{center}
\caption{The density of eigenvalues $\rho_{\ec}(\lambda)$ 
for exponential matrix $\A$, $\C=\1_N,r=0.2$ and for three 
different autocorrelation times $t$: $t=0$ (solid line), 
$t=2$ (dashed line) and $t=5$ (dotted line).}
\label{fig1}
\end{figure}

Below we will discuss the case of exponential autocorrelations.  
Exponential correlations are encountered in many situations. 
The general solution, which we have discussed so far, simplifies
in this case to a more compact relation for the Green's function,
which allows us to find analytically an approximate form of the
eigenvalue density of the random matrices $\ec$ and $\ea$.
The approximation becomes exact in the large $N$ limit. 
We consider purely exponential autocorrelations given by
the autocorrelation matrix:
\begin{equation}
A_{\alpha\beta}=\exp \left[ -|\alpha\!-\!\beta|/t \right] \ ,
\label{Aexp}
\end{equation}
where $t$ controls the range of autocorrelations.
The inverse of the matrix ${\A}$ reads:
\begin{equation}
{\A}^{-1} = \frac{1}{2 sh } \left[\begin{array}{rrrrrr} 
ex,  & -1, &        &&& \\
-1, & 2 ch, & -1,     &&& \\
   &    & \dots & \dots & \dots & \\ 
   &    &    & -1, & 2 ch, & -1  \\
   &    &    &    & -1, & ex \end{array} \right].
\label{Binv}
\end{equation}
We have introduced here a shorthand notation 
$ex=\exp(1/t)$, $ch = \cosh(1/t)$ and $sh = \sinh (1/t)$.
The spectrum of this matrix can be 
approximated by the spectrum of a matrix $\M$:
\begin{equation}
\M = \frac{1}{2 sh } \left[\begin{array}{rrrrrr} 
2ch,  & -1, &        &&& -1 \\
-1, & 2 ch, & -1,     &&& \\
   &    & \dots & \dots & \dots & \\ 
   &    &    & -1, & 2 ch, & -1  \\
-1,  &    &    &    & -1, & 2ch \end{array} \right],
\label{M}
\end{equation}
whose eigenvalues can be found analytically:
\begin{eqnarray}
\mu_\alpha = (ch + \cos(\pi \alpha/T))/sh.
\nonumber
\end{eqnarray}
The corresponding eigenvectors are given by the Fourier
modes. The matrix $2 sh \cdot \M$ can be viewed as of a sum:
$(2ch - 2) \1_T + \Delta_T$, of
a unity matrix multiplied by a constant and a
discretized one-dimensional Laplacian $\Delta_T$
for a cyclic chain of length $T$. The matrix ${\A}^{-1}$ 
can be obtained from $\M$ by adding to it a perturbation 
$\mathbf{P}$: ${\A}^{-1}=\M+\mathbf{P}$, 
where $\mathbf{P}$ has only four non-vanishing elements: 
$P_{11}=P_{TT} = ex^{-1}$ and $P_{1T}=P_{T1}=1$.
The first order corrections to the eigenvalues of ${\A}^{-1}$,
which stem from the perturbation $\mathbf{P}$, behave as $1/T$. 
The perturbation $\mathbf{P}$ can be viewed as a change of a boundary
condition of the Laplacian. As usual, boundary conditions
affect mostly the longest (small momentum) modes.
Indeed, a careful analysis shows that the two diagonal terms 
of the perturbation matrix, $P_{11}=P_{TT}$, introduce a constant
correction independent of $T$ of the lowest eigenvalues which 
does not vanish when $T$ goes to infinity.
However, since the differences between unperturbed eigenvalues
of $\M$ and the corresponding perturbed eigenvalues of ${\A}^{-1}$
disappear for all other eigenvalues, 
we expect that for $t\ll T$ the spectral properties of ${\A}$ 
can be well approximated by the eigenvalues of $\M^{-1}$:
\begin{equation}
\Lambda_\alpha \approx \frac{1}{\mu_\alpha} = 
\frac{sh }{ch + \cos (\pi \alpha/T)} \ .
\end{equation}
In this limit we can also approximate the 
sum (\ref{zZsum}) by an integral:
\begin{equation}
\frac{z}{Z} \approx \frac{sh}{\pi}\int\limits_0^{\pi} 
\frac{d\tau}{(ch - sh \cdot F) + \cos{\tau}} \ ,
\label{zZint}
\end{equation}
Where $\tau = \pi \alpha/T$ and the symbol 
$F \equiv \frac{Z}{z} r m_{\ec}(z)$ is introduced for brevity. Note that
in the definition of $F$ we have replaced $M_{\C}(Z)$,
which would rather be dictated by (\ref{zZsum}), by
$m_{\ec}(z)$. This change is legitimate due to (\ref{map}). 
The integral (\ref{zZint}) can be done:
\begin{equation}
\frac{z}{Z} = \frac{sh}{\sqrt{(ch-sh\,F)^2-1}} \ .
\end{equation}
Setting back $F= \frac{Z}{z} r m_{\ec}(z)$ we eventually obtain:
\begin{equation}
Z = z \frac{ch \cdot r m_{\ec}(z) - \sqrt{sh^2 + r^2 m^2_{\ec}(z)}}{
sh \cdot (r^2 m^2_{\ec}(z) -1)} \ .
\label{Z(z)}
\end{equation}
This is an explicit equation for $Z = Z(z)$ which can be now
inserted into $m_{\ec}(z)=M_{\C}(Z)$ giving us a compact
relation for $m_{\ec}(z)$ 
in the presence of the exponential autocorrelations (\ref{Aexp}):
\begin{equation}
m_{\ec}(z) = 
M_{\C}\left(
 z \frac{ch \cdot r m_{\ec}(z) - \sqrt{sh^2 + r^2 m^2_{\ec}(z)}}{
sh \cdot (r^2 m^2_{\ec}(z) -1)} 
\right) \ .
\label{mMexp}
\end{equation}
In the limit $t \rightarrow 0$, the parameters 
$sh=\sinh (1/t)$, $ch=\cosh (1/t)$ and $ex =\exp (1/t)$ increase
to infinity and $ch/ex \approx sh/ex \approx 1$. As a consequence, 
the form of equation (\ref{Z(z)}) simplifies to (\ref{Zz}), which
corresponds to the case without autocorrelations, as expected.
Using the equation (\ref{mMexp}) we can recursively generate 
equations for the consecuitive moments: 
\begin{eqnarray}
&& m_{{\ec}1} = M_{{\C}1}  \nonumber \\
&& m_{{\ec}2}  = M_{{\C}2} + rM_{{\C}1}^2 \cdot cth  \nonumber \\
&& m_{{\ec}3}  = M_{{\C}3} + 3rM_{{\C}1}M_{{\C}2} \cdot cth + 
r^2 M_{{\C}1}^3 \left(\frac{3}{2} \; cth^2 - \frac{1}{2}\right) 
\label{mMsc} \\
&& m_{{\ec}4} = M_{{\C}4} + 
2 r (M_{{\C}2}^2 + 2 M_{{\C}1} M_{{\C}3}) \cdot cth
+ 2 r^2 M_{{\C}1}^2 M_{{\C}2} \left( 4 cth^2 -1\right) + 
r^3 M_{{\C}1}^4 \left( \frac{5}{2} cth^3 - \frac{3}{2} cth\right)
\nonumber \\
      &&  \dots \ , \nonumber
\end{eqnarray}
where $cth=ch/sh =\coth(1/t)$.
The coefficients on the right hand side, which depend
on $cth$, can be directly expressed in terms of
the moments $M_{{\A}k}$ of the matrix ${\A}$. 
Approximating again a sum by an integral 
in the large $T$ limit we can write:
\begin{eqnarray}
M_{{\A}k} = \frac{1}{T} \sum\limits_{\alpha=1}^{T} \Lambda_\alpha^k
\approx \frac{sh^k}{\pi} 
\int\limits_0^{\pi} \frac{d\tau}{(ch+ \cos{\tau})^k} \ .
\nonumber
\end{eqnarray}
The integrals can be calculated yielding:
\begin{eqnarray}
\begin{array}{rl}
M_{{\A}1} & = 1 \\
M_{{\A}2} & = cth \\
M_{{\A}3} & = \frac{3}{2} cth^2 - \frac{1}{2} \\ \\
M_{{\A}4} & = \frac{5}{2} cth^3 - \frac{3}{2} cth \\
          & \dots 
\end{array}
\label{approx}
\end{eqnarray}
We see that if we insert these coefficients into the equations
(\ref{mcMM}) we obtain (\ref{mMsc}). This is a consistency
check for the approximation which we use here. 
The quality of this approximation can also be checked
by comparing the moments $M_{{\A}k}$ of the matrix
(\ref{Aexp}) for finite $T$ with the result (\ref{approx})
which corresponds to $T=\infty$. We expect that for $t\ll T$
the numerical values shall approach the result (\ref{approx}).
The results of this comparison confirm our expectations (see table 1).
\begin{table}
$$
\begin{array}{| r | l | l | l |}
\hline
\multicolumn{4}{|c|}{t=1} \\
\hline
T  & M_{{\A}2} & M_{{\A}3} & M_{{\A}4} \\
\hline
 20    & 1.29493   & 2.01479   & 3.48620  \\
 50    & 1.30579   & 2.05757   & 3.60838  \\
100    & 1.30941   & 2.07183   & 3.64911  \\
200    & 1.31123   & 2.07896   & 3.66947  \\
500    & 1.31231   & 2.08324   & 3.68169  \\
\hline 
\infty & 1.31304   & 2.08609   & 3.68983  \\
\hline 
\end{array}
\qquad
\begin{array}{| r | l | l | l |}
\hline
\multicolumn{4}{|c|}{ t=5}\\
\hline
T  & M_{{\A}2} & M_{{\A}3} & M_{{\A}4} \\
\hline
 20    & 4.44996   & 28.6455   & 204.107  \\
 50    & 4.81980   & 34.2544   & 271.932  \\
100    & 4.94314   & 36.1292   & 294.733  \\
200    & 5.00482   & 37.0666   & 306.133  \\
500    & 5.04182   & 37.6290   & 312.973  \\
\hline 
\infty & 5.06649   & 38.0040   & 317.534  \\
\hline 
\end{array}
\qquad
\begin{array}{| r | l | l | l |}
\hline
\multicolumn{4}{|c|}{ t=10}\\
\hline
T  & M_{{\A}2} & M_{{\A}3} & M_{{\A}4} \\
\hline
 20    & 7.58726   & 79.6134   & 891.336  \\
 50    & 9.03668   & 120.509   & 1784.56  \\
100    & 9.53497   & 135.501   & 2146.93  \\
200    & 9.78414   & 143.001   & 2328.48  \\
500    & 9.93364   & 147.501   & 2437.40  \\
\hline 
\infty & 10.0333   & 150.501   & 2510.02  \\
\hline 
\end{array}
$$
\caption{The moments $M_{{\A}2}$, $M_{{\A}3}$ and $M_{{\A}4}$
of the matrix $\A$ (\ref{Aexp}) for three values of the
autocorrelation length $t=1,5,10$ calculated numerically
for finite size $T=20,\dots,500$ and by the analytic formula
(\ref{approx}) which corresponds to $T=\infty$. The finite size
values approach the values given by (\ref{approx}) as $T/t$ tends
to infinity.}
\end{table}
Thus we see that the formula (\ref{mMsc}) for $m_{\ec}(z)$
in the presence of the exponential autocorrelations 
becomes exact in the limit $T\rightarrow \infty$.
This formula allows us to compute the eigenvalue distribution
of the random matrices $\ec$ and $\ea$ (\ref{ca}). Let us illustrate
this on the simplest example of the system which has no 
correlations: $\C=\1_N$ and 
$M_{\C}(Z) = 1/(Z-1)$. 
In this case the equation (\ref{mMexp}) for $m_{\ec}(z)$
takes the form:
\begin{equation}
cth \cdot x^2 - \frac{1}{z}(x^2-1)(x+r) - x\sqrt{1+x^2/sh^2} = 0 \ ,
\label{eq:x}	
\end{equation}
where we have used the notation $x(z) = r m_{\ec}(z)$.
For $t\rightarrow 0$ ($\A\rightarrow \1_T$) this equation 
reduces to:
\begin{equation}
(x-1)\left[x-\frac{1}{z}(x+1)(x+r)\right] = 0 \ ,
\end{equation}
which has a solution:
\begin{equation}
x(z) = \frac{1}{2}\left(-1-r-i\sqrt{(\mu_{+}-z)(z-\mu_{-})}+z\right),
\end{equation}
where $\mu_{\pm} = (1\pm\sqrt{r})^2$, which leads to the well-known 
result for the uncorrelated Wishart ensemble:
\begin{equation}
\rho_{\ec}(\lambda) = 
\frac{1}{2\pi r} \frac{\sqrt{(\mu_+-\lambda)(\lambda -\mu_-)}}{\lambda} \ .
\end{equation}
For $t>0$ it is still possible to find analytically 
a solution of (\ref{eq:x}). 
Let us rewrite (\ref{eq:x}) as a polynomial equation:
\begin{eqnarray}
-z^2 x^2 (1+x^2/sh^2 ) + 
\left(cth\,zx^2 - (r+x)(x^2-1)\right)^2 = 0 \ . \nonumber
\end{eqnarray}
It has two trivial solutions $x=\pm1$. Dividing out
the polynomial $(x-1)(x+1)$ we get:
\begin{equation}
x^4 + 2x^3(r - cth\,z)  + 
x^2(-1 + r^2 - 2cth\,rz + z^2) - 2rx -r^2 = 0 \ .	\label{poly4}
\end{equation}
This is a quartic equation which can be solved analytically
by the Ferrari method. We will not present the formal solution 
which is neither transparent nor informative.
Instead, we show in Fig. \ref{fig1} 
the eigenvalue density functions $\rho_{\ec}(\lambda)$, for
different $t$, resulting from this solution. 
The lower part of the distribution 
approaches zero when $t$ increases, but zero modes do
not appear in the distribution as long as $r<1$.
The formula (\ref{mMexp}) applies to any correlation
matrix $\C$ but in the general case one has to use a numerical
procedure to calculate from it the density function. 

Let us stop here the presentation of results for
the ensemble of real matrices. As we
mentioned all results in the large $N$ limit hold also in
the ensemble of complex matrices if the covariance matrices
(\ref{ca}) are replaced by (\ref{cac}). The reason why it is
so is related to the fact that
the moments of $\ec = \frac{1}{T} \X \X^\dagger$
in the ensemble of complex matrices (\ref{pcompl})
are equal to
the moments of $\ec = \frac{1}{T} \X \X^\tau$
in the real ensemble (\ref{preal}) up to a $1/N$ corrections
which disappear in the large $N$ limit:
\begin{equation}
\frac{1}{N} \left\langle \left(\frac{1}{T} \X \X^\dagger \right)^k
\right\rangle_{\mbox{\small complex}} =
\frac{1}{N}\left\langle \left(\frac{1}{T} \X \X^\tau \right)^k 
\right\rangle_{\mbox{\small real}} + O(1/N)
\end{equation}
Let us illustrate this by explicit calculations
of the second moment. Using the Wick's theorem for Gaussian
integrals and the equation (\ref{CA}) for the two-point correlation
function, we have:
\begin{eqnarray}
&& \frac{1}{N} \left\langle \left(\frac{1}{T} \X \X^\tau \right)^2 
\right\rangle = 
\frac{1}{NT^2}
\left\langle X_{i\alpha} X_{j\alpha} X_{j\beta} X_{i\beta} 
\right\rangle = \nonumber \\
&& = \frac{1}{NT^2} \left\{
\left\langle X_{i\alpha} X_{j\alpha} \right\rangle 
\left\langle X_{j\beta} X_{i\beta} \right\rangle +
\left\langle X_{i\alpha} X_{i\beta} \right\rangle 
\left\langle X_{j\alpha} X_{j\beta} \right\rangle +
\left\langle X_{i\alpha} X_{j\beta} \right\rangle
\left\langle X_{j\alpha} X_{i\beta} \right\rangle \right\} \nonumber \\
&& = \frac{1}{NT^2} \left\{ C_{ij} A_{\alpha\alpha} C_{ji} A_{\beta\beta} +
                  C_{ii} A_{\alpha\beta} C_{jj} A_{\alpha\beta} +
                  C_{ij} A_{\alpha\beta} C_{ji} A_{\alpha\beta} \nonumber 
\right\} \\
&& = M_{{\C}2} M_{{\A}1}^2 + 
     r M_{{\C}1}^2 M_{{\A}2} + \frac{r}{N} M_{{\C}2} M_{{\A}2} \ . \nonumber 
\end{eqnarray}
The corresponding calculations
for the complex ensemble read:
\begin{eqnarray}
&& \frac{1}{N} \left\langle \left(\frac{1}{T} \X \X^\dagger \right)^2 
\right\rangle = 
\frac{1}{NT^2}
\left\langle X_{i\alpha} X^*_{j\alpha} X_{j\beta} X^*_{i\beta} 
\right\rangle = \nonumber \\
&& = \frac{1}{NT^2} \left\{
\left\langle X_{i\alpha} X^*_{j\alpha} \right\rangle 
\left\langle X_{j\beta} X^*_{i\beta} \right\rangle +
\left\langle X_{i\alpha} X^*_{i\beta} \right\rangle 
\left\langle X_{j\alpha} X^*_{j\beta} \right\rangle +
\left\langle X_{i\alpha} X_{j\beta} \right\rangle
\left\langle X^*_{j\alpha} X^*_{i\beta} \right\rangle \right\} \nonumber \\
&& = \frac{1}{NT^2} \left\{ C_{ij} A_{\alpha\alpha} C_{ji} A_{\beta\beta} +
                  C_{ii} A_{\alpha\beta} C_{jj} A_{\alpha\beta} \right\}
                  \nonumber \\
&& = M_{{\C}2} M_{{\A}1}^2 + 
     r M_{{\C}1}^2 M_{{\A}2} \ . \nonumber 
\end{eqnarray}
The difference between the two calculations appears in
the third term which in the real ensemble gives a contribution 
of the order $1/N$ while in the complex ensemble disappears by virtue 
of (\ref{CAcc}). We recognize that
$M_{{\C}2} M_{{\A}1}^2 + r M_{{\C}1}^2 M_{{\A}2}$  
which are the leading terms in the $1/N$ expansion are identical as in the
second equation in the set (\ref{mcMM}). Generally one can show
that the leading contributions which correspond to the planar
diagrams in the expansion of $\gc(z)$ are identical for both
ensembles. Non-planar diagrams are different but they contribute
in the subleading orders: it turns out that in the diagrammatic 
expansion of the Green's function $g_c(z)$ for the complex matrix
ensemble (\ref{pcompl}), which would be a counterpart of 
(\ref{eq:fgraph}) in the appendix, all diagrams which contain
a double arc with dashed and solid line crossed 
are identically equal zero since such an
arc corresponds to the propagator 
$\langle X_{i\alpha} X_{j\beta} \rangle$ or
$\langle X^*_{i\alpha} X^*_{j\beta} \rangle$.
A crossing of two arcs is however allowed and leads 
to a factor $1/N^2$.

To summarize: in the paper we have considered an Wishart ensemble of 
correlated random matrices. 
We have obtained in the limit of large matrices a closed set of equations
relating the Green's function or equivalently the moments'
generating functions $m_{\ec}(z)$ and $m_{\ea}(z)$ for statistically
dressed correlations to the generating functions for genuine
correlation matrices $M_{\C}(z)$ and $M_{\A}(z)$. The equations in
the large $N$ limit are the same for the ensemble of real and complex matrices.
Using these equations we can write down exact relations between genuine and 
experimental spectral moments of correlation functions of an arbitrary
order. The relations can be used in practical problems to learn about
correlations in the studied system from the experimental samples.
In the case of exponential correlations we have also found an explicit
form the spectral density function of the covariance matrix.
A natural generalization of the work presented here is to consider
a more general type of time correlations than purely
exponential (\ref{Aexp}). If the correlations are of the form which
depends on the time difference $A_{\alpha,\beta} = A(|\alpha-\beta|)$
and if they are short-ranged
then one can apply Fourier transform to determine 
in the large $T$ limit an approximate spectrum of the matrix $\A$ 
and approximate values of its spectral moments. Another interesting
issue which can be addressed in the future is the determination of the
probability distribution for individual elements of the covariance 
matrices $\ec$ and $\ea$ similarly as it was done for the
uncorrelated case \cite{jn}.

\bigskip

\noindent
{\bf Acknowledgments}

\medskip

\noindent
We thank R. Janik, A. Jarosz and M.A. Nowak for discussions.
This work was partially supported by 
the Polish State Committee for
Scientific Research (KBN) grants
2P03B 09622 (2002-2004) and 2P03B-08225 (2003-2006),
and by EU IST Center of Excellence "COPIRA".

\section*{Appendix 1}
For completeness we recall here the graphical
representation of the Green's function. The details
of the diagrammatic method can be found in \cite{fz,sm3,bgjj}.
The Green's function $\gc(z)$ can be represented as
a sum over diagramms:
\begin{equation}
\psfrag{+}{$+$}\psfrag{=}{$=$}\psfrag{g}{${\gc}$}\psfrag{k}{$\dots$}
\includegraphics[width=11cm]{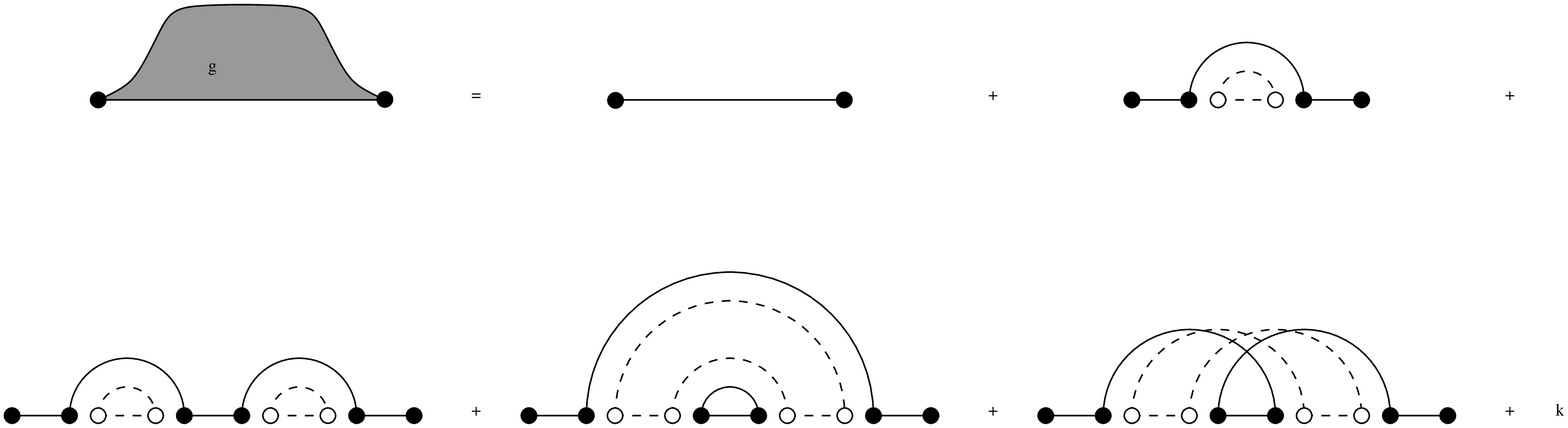}
\label{eq:fgraph}
\end{equation}
where the $N$-type and $T$-type indices of $\X$ are denoted 
by filled and empty circles, respectively. 
The matrix $\X$ is denoted by an ordered pair of neighbouring 
filled and empty circles, while $\X^{\tau}$ is drawn as an pair 
of such circles in the reverse order. 
A horizontal solid line stands for $\1_N/z$, a dashed line for
$\1_T/T$, a solid arc for $\C$ and a dashed arc for $\A$. 
The two point function (\ref{CA}) is drawn as a double arc.
Matrices on a line are multiplied in the order of appearance on 
this line. If a line is closed, the trace is taken.

In the thermodynamical limit only planar diagrams give 
contribution to $\gc$. In particular the last term in
(\ref{eq:fgraph}) vanishes. The Green's function $\gc_*(z)$ 
is represented by an identical set of diagrams with dashed
and solid lines exchanged. It is convenient to introduce
one-line irreducible diagrams and corresponding generating
functions $\Sigmac$ and $\Sigmacc$. The
Green's functions can be expressed in terms of
$\Sigmac$ and $\Sigmacc$ as follows:
\begin{equation}
\psfrag{s}{{\small $\Sigmac$}}\psfrag{g}{{\small $\gc$}}
\psfrag{k}{{\small $\dots$}}
\includegraphics[width=13cm]{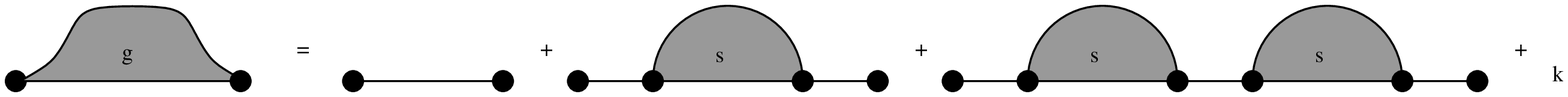} 
\end{equation}
\begin{equation}
\psfrag{s}{{\small $\Sigmacc$}}
\psfrag{g}{{\small $\gcc$}}\psfrag{k}{{\small $\dots$}}
\includegraphics[width=13cm]{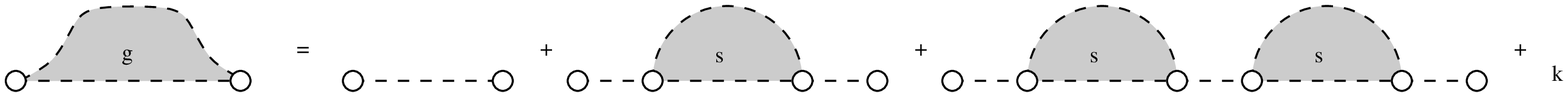}
\end{equation}
In the planar limit there are two additional equations
which relate the sums over one-line irreducible diagrams to
the Green's functions:
\begin{equation}
\psfrag{s}{{\small $\Sigmacc$}}\psfrag{g}{{\small $\gc$}}
\psfrag{k}{{\small $\dots$}}
\includegraphics[width=4.5cm]{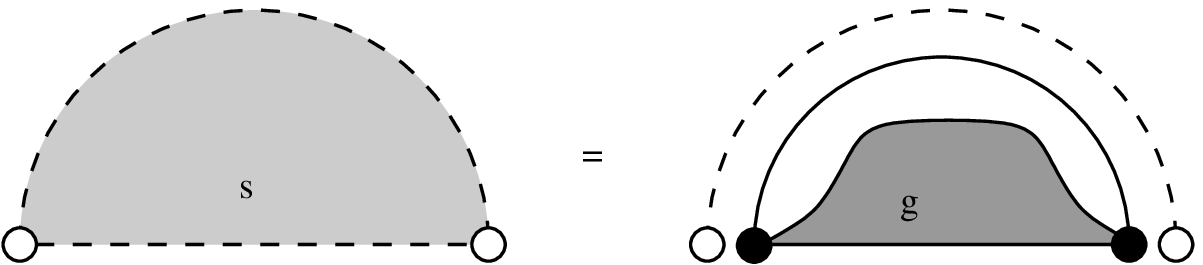} \qquad , \qquad
\psfrag{s}{{\small $\Sigmac$}}\psfrag{g}{{\small $\gcc$}}
\psfrag{k}{{\small $\dots$}}
\includegraphics[width=4.5cm]{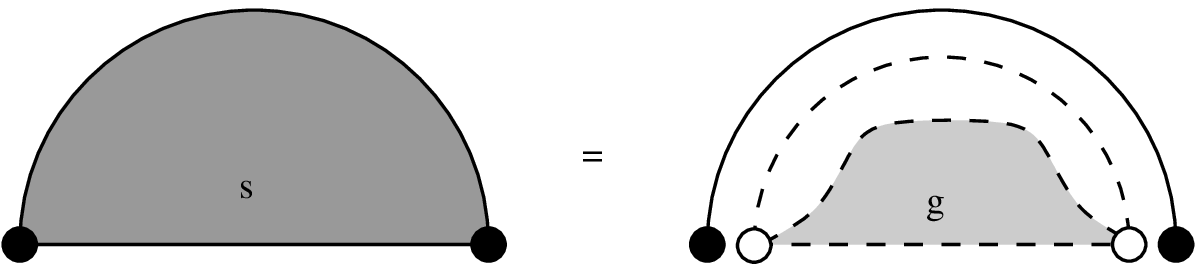}
\end{equation}
Analogous diagrammatic equations can be written
for $\mathbf{g_a}$ with the only difference that the solid 
line shall denote the propagator $\1_T/z$ and the dashed line
$\1_N/N$. 

\section*{Appendix 2}
We use the equations (\ref{zZ_CA}) and (\ref{map})
to determine the relations (\ref{mcMM}).
As for the case $\A=\1_T$ we shall do this using $1/z$ expansion.
The function $M_{\A}(Z)$ is given by the series:
\begin{equation}
M_{\A}(Z) = \frac{M_{{\A}1}}{Z} + \frac{M_{{\A}2}}{Z^2} + 
\frac{M_{{\A}3}}{Z^3} + \dots \ .
\end{equation}
Let us determine the expansion
for the inverse function $M_{\A}^{-1}$
as a series around zero:
\begin{equation}
M^{-1}_{\A}(y) = 
M_{\A 1} y^{-1} \left( 1 + \mu_1 y + \mu_2 y^2 + \dots\right).
\end{equation}
The coefficients of the series can be directly calculated 
from the condition:
\begin{equation}
y=M_{\A}(M_{\A}^{-1}(y)) \ ,
\end{equation}
which gives us:
\begin{equation}
\mu_1 = \frac{M_{\A 2}}{(M_{\A 1})^2} \quad , \quad
\mu_2 = \frac{M_{\A 3} M_{\A 1} - 
(M_{\A 2})^2}{(M_{\A 1})^4} \quad , \quad \dots \quad .
\end{equation}
The equation (\ref{zZ_CA}) takes the form:
\begin{equation}
z = M_{{\A}1} Z \left( 1 + \mu_1 r M_{\C}(Z) + 
\mu_2 r^2 M^2_{\C}(Z) + \dots \right) \ , \label{z=...}
\end{equation}
or if written for $1/z$:
\begin{equation}
\frac{1}{z} = \frac{1}{M_{{\A}1}} \frac{1}{Z} 
\left( 1 - \mu_1 r M_{\C}(Z) + 
(\mu^2_1 - \mu_2) r^2 M^2_{\C}(Z) + \dots \right) \label{1/z=...},
\end{equation}
where:
\begin{equation}
M_{\C}(Z) = \frac{M_{{\C}1}}{Z} + \frac{M_{{\C}2}}{Z^2} + 
\frac{M_{{\C}3}}{Z^3} + \dots \ .
\end{equation}
Thus we have expressed $1/z$ as a series of $1/Z$.
Inserting this series to the equation (\ref{map}) and comparing
coefficients at $1/Z^k$ we eventually obtain (\ref{mcMM}).


\begin{thebibliography}{99}
\bibitem{ret} Because of the multiplicative nature of 
the price changes instead of the changes:
$X_{it} = P_{it+1} \! - P_{it}$ themselves,
in the financial analysis one rather uses 
returns $X_{it} = \log P_{it+1}/P_{it}$ as random variables.
\end{thebibliography}
\end{document}